# An X-ray Synchrotron Shell and a Pulsar: The Peculiar Supernova Remnant G32.4+0.1

Stephen P. Reynolds [1] and Kazimierz J. Borkowski [1]

[1]*Physics Department*
*North Carolina State University*
*PO Box 8202 Raleigh, NC 27695-8202, USA*

## ABSTRACT

We present a deep *Chandra* observation of the shell supernova remnant G32.4+0.1, whose featureless X-ray spectrum has led to its classification as an X-ray synchrotron-dominated supernova remnant (SNR). We find a partial shell morphology whose outline is quite circular, with a radius of about 11 pc at an assumed distance of 11 kpc. Thermal and power-law spectral models for three relatively bright regions provided equally good fits, but the absence of spectral lines required ionization timescales from thermal fits that are inconsistent with mean densities derived from emission measures. We thus confirm the nonthermal, i.e., synchrotron, origin of X-rays from G32.4+0.1. Shock velocities needed to accelerate electrons to the required TeV energies are $\gtrsim 1000$ km s$^{-1}$, giving remnant ages $\lesssim 5{,}000$ – 9,000 yr. There is no obvious X-ray counterpart to the radio pulsar PSR J1850−0026, but its position adjoins a region of X-ray emission whose spectrum is somewhat harder than that of other regions of the shell, and which may be a pulsar-wind nebula (PWN), though its spectrum is steeper than almost all known X-ray PWNe. The distance of the pulsar from the center of symmetry of the shell disfavors a birth in a supernova event at that location only a few thousand years before: either the pulsar (and putative PWN) are not associated with the shell SNR, requiring a coincidence of both position and (roughly) absorbing column density, or the SNR is much older, making the origin of nonthermal emission problematic.

*Keywords:* Supernova remnants (1667), pulsars (1306), pulsar-wind nebulae (2215)

## 1. INTRODUCTION

Great progress has been made in the last decade or two in understanding the nature of particle acceleration at strong shock waves. Nonthermal particle distributions are found in virtually all astrophysical environments, from molecular clouds where the ionization they produce is central to the physics of star formation, to clusters of galaxies hosting puzzling radio relics and halos, so understanding how and where particles are accelerated is a fundamental astrophysical problem of wide importance. A major part of the recent progress is owed to the study of Galactic supernova remnants (SNRs): prime laboratories where the impulsive injection of energy and resulting shock waves can be studied with both spatial and spectral resolution, and where the local conditions – gas density, ionization state, shock speeds – can be well characterized.

Nonthermal radiation from SNRs is now well-established to extend to X-ray wavelengths, requiring electrons with TeV energies if the emission process is synchrotron radiation, as is widely accepted (see arguments in, e.g., Reynolds 2008), and which has now been conclusively demonstrated by the detection of polarized X-rays in Cas A (Vink et al. 2022) and Tycho (Ferrazzoli et al. 2023) with the X-ray polarimetry mission IXPE. While it is now known that most SNRs younger than a few thousand years show evidence for a synchrotron component in addition to thermal X-ray emission (e.g., Bamba et al. 2005; Acero et al. 2016), the synchrotron contribution is often small and hard to distinguish from thermal emission. For a few SNRs, however, synchrotron X-rays dominate, as in the original Rosetta Stone object SN 1006 (Reynolds & Chevalier 1981; Koyama et al. 1995), simplifying the use of X-ray information to constrain the physics of electron acceleration. See Reynolds (2008) for a review, and Acero et al. (2016) for a recent list of X-ray synchrotron SNRs (XSSNRs). In these objects, the roughly power-law electron energy distribution extending from radio wavelengths is observed to have steepened, in a gradual rolloff due either to synchrotron losses or to the finite time available for electron acceleration. This steepening contains information on the maximum energy to which electrons have been accelerated, which in turn encodes information on the acceleration process. The sample size remains small, however, so each additional object can contribute important information. Here we report *Chandra* observations of one lesser-known object, G32.4+0.1, claimed to be a member of this class (Yamaguchi et al. 2004), which, uniquely among the XSSNR sample, also contains a radio pulsar within its boundaries. Section 2 describes G32.4+0.1 in some detail. Section 3 summarizes the observations. Sections 4 and 5 exhibit imaging and spectroscopic results, respectively. Sections 6 and 7 discuss and summarize our results.

## 2. G32.4+0.1

This remnant, G32.4+0.1 (also called G32.45+0.1), was first identified as a hard-spectrum, apparently lineless source from the *ASCA* Galactic Plane Survey (Yamauchi et al. 2002). A short XMM-*Newton* observation of G32.4+0.1, about 21 ks (after filtering) for each of MOS1 and MOS2, was performed in 2003 (Yamaguchi et al. 2004). The image from that observation shows an irregular X-ray shell of radius about 4′. Yamaguchi et al. (2004) report a highly absorbed spectrum ($N_H \sim 5 \times 10^{22}$ cm$^{-2}$) well described by a power-law with photon index $\Gamma = 2.2$ (1.8 – 3.0) (90% confidence interval, as are all confidence intervals quoted in this paper, except where otherwise noted). This fit gives an absorption-corrected $0.5 - 10$ keV flux of $2.8 \times 10^{-12}$ erg cm$^{-2}$ s$^{-1}$. A thermal model (NEI) fit was only slightly worse statistically, but no lines are apparent, and the electron temperature from that fit of about 5 keV is highly implausible.

The pulsar, PSR J1850-0026, was discovered by Keith et al. (2009) in the Parkes Multibeam Pulsar Survey. It is located off-center by about 2.5′, but well within the diffuse X-ray emission of the shell. The pulsar is undistinguished, with a period of 166 ms, a spindown luminosity $\dot{E}$ of $3.3 \times 10^{35}$ erg s$^{-1}$ and a characteristic age of 68 kyr (Keith et al. 2009). Of course, the true age could be much younger if the current period is close to the birth period.

The distance to the pulsar is based on the observed dispersion measure of 947 cm$^{-3}$ pc (Keith et al. 2009), who use two different Galactic electron-density models to obtain distances of 10.8 and 11.2 kpc. Various distances $d$ to the shell have been quoted; Yamaguchi et al. (2004) divided their $N_H$ value of $5.2 \times 10^{22}$ cm$^{-2}$ by an assumed mean ISM hydrogen density of 1 cm$^{-3}$ to obtain $d = 17$ kpc, while Kilpatrick et al. (2016) used the velocity of broadened CO J = 2-1 emission ($\sim +45$ km s$^{-1}$) seen toward the eastern shell of G32.4+0.1 to obtain $11.8 - 18.5$ kpc from rotation curves. A cavity in $^{12}$CO (1-0) emission at a velocity of 10.8 km s$^{-1}$ surrounding G32.4+0.1 was identified in a survey of first-quadrant SNRs (Sofue et al. 2021), but a survey by Zhou et al. (2023) found broad CO (1-0) line features associated with G32.4+0.1 at a velocity giving $d = 10.1 \pm 0.4$ kpc. Though an association of the shell and the pulsar is subject to question (see below), we shall use the pulsar distance of 11 kpc as the distance to the shell. The remnant radius is then about 11 pc.

A shell-like radio source coincident with the X-ray shell was identified by Yamaguchi et al. (2004) from the NRAO VLA Sky Survey at 1.4 GHz (Condon et al. 1998), with a resolution of 45″ FWHM. The source was further studied by Dokara et al. (2023) as part of the GLOSTAR Galactic-plane survey (Brunthaler et al. 2021). Dokara et al. (2023) determined a flux at 1.4 GHz of $0.75 \pm 0.13$ Jy, and a spectral index of $\alpha = -0.39 \pm 0.1$ from a spatially resolved TT analysis. The 5.9 GHz image shows substantial emission in filamentary structures away from the X-ray shell, which we interpret as unrelated thermal emission, whose flat spectrum would account for its not being obvious at 1.4 GHz.

G32.4+0.1 is listed as a "marginally classified candidate" in the first Fermi/LAT SNR catalog (Acero et al. 2016), with the marginal status based partly on confusion due to possible interstellar emission. The nominal fitted parameters are a photon flux of $4.8 \times 10^{-9}$ ph cm$^{-2}$ s$^{-1}$ between 1 and 100 GeV, and an unusually steep photon index $\Gamma = 4.1 \pm 0.6$ – the steepest of any source listed in the Acero et al. (2016) catalog. No TeV counterpart has been reported. The GeV flux is unlikely to be associated with either the pulsar or the shell, for a distance of 11 kpc; at that distance, it roughly corresponds to a luminosity of $1 \times 10^{35}$ erg s$^{-1}$, unlikely to be powered by a pulsar with $\dot{E}$ only about three times larger. The collection of Grenier & Harding (2015) shows only one (of several dozen) rotation-powered pulsar with a luminosity above 100 MeV comparable to its spindown luminosity. No GeV source that luminous (attributable either to the pulsar or a PWN) was found in the Fermi-LAT survey of TeV-emitting PWNe (Acero et al. 2013), for log $\dot{E} = 35.5$. It would also be exceptional for a SNR, larger than all but two of the 25 GeV SNRs catalogued by Acero et al. (2016) (along with a larger value of $\Gamma$ than any of those).

Fundamental questions still surround the nature of G32.4+0.1: can the X-ray spectrum be more confidently identified as nonthermal, and if not, what are the thermal properties? Does the pulsar, known to date only in radio, have an X-ray counterpart? Is the pulsar really associated with the shell? Might we identify a pulsar-wind nebula powered by this relatively young pulsar? We obtained a long *Chandra* observation to address these questions.

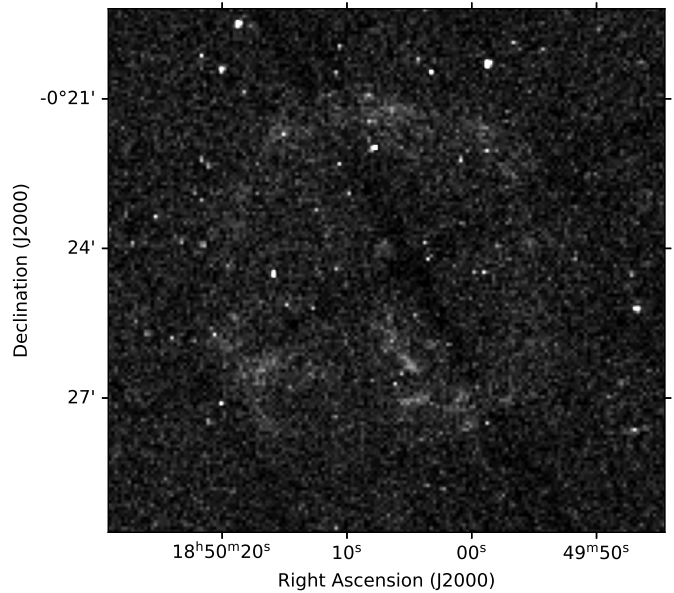

**Figure 1.** Raw counts image of G32.4+0.1, between 1.5 and 6.5 keV, binned by a factor of 8 (3.94″ pixels). Scale is linear up to 20 counts per (binned) pixel and is cut off above.

## 3. OBSERVATIONS

We observed G32.4+0.1 for a total of 182.9 ks in eight segments (ObsIDs 22991, 22992, 23345, 23346, 23357, and 22429 between 31 July and 10 August 2020, and ObsIDs 22430 and 25100 on 5 and 6 August 2021), in VFAINT mode, using the ACIS-I array. We reprocessed the data using CIAO version 4.15 and CALDB version 4.10.4. The individual segments were merged using observation 22991 as reference. The slightly different roll angles of the eight observations required us to extract all spectra individually from each observation; these were then combined, along with the responses, with appropriate weightings.

## 4. IMAGING

Figures 1 and 2 show raw (but binned by factors of 8 and 4, respectively) and smoothed images of G32.4+0.1, in the energy range of 1.5 to 6.5 keV, since outside that range, no source emission is evident. In order to show faint structure, all images are linear with a cutoff above a maximum count rate. Since a significant fraction of the background is due to particle interactions (see below), we have not attempted an exposure correction, which should not apply to the particle background. The source is extremely faint, with a mean surface brightness between 1.5 and 6.5 keV of only 0.045 raw counts/pixel within a circular region of radius 210″, shown in yellow on Figure 2. While the brightness is uneven, emission appears to outline a clear fairly circular shell. Three brighter regions, "Northwest shell," "South interior," and "East", as indicated on Figure 2 left, were selected for spectral analysis below, and have mean surface brightnesses of 0.055 to 0.072 counts/pixel. (Background regions have typically 0.03 counts/pixel.)

Figure 3, left, shows an RGB image of 1.4 GHz radio emission from the NRAO VLA Sky Survey (Condon et al. 1998) in red, 5.9 GHz radio emission (Dokara et al. 2023) in green, and X-rays in blue. Both 5.9 GHz and X-ray images have been smoothed to approximate the 1.4 GHz image resolution; the X-ray image of Figure 1, with 3.94″ pixels, was smoothed with a Gaussian of $\sigma = 9$ pixels. Correspondence is relatively good in the west and southwest of the shell, but radio emission falls interior to X-rays in the east. Figure 4 shows azimuthally averaged radial profiles of the 5.9 GHz radio image and the X-ray image. For brighter synchrotron-dominated SNRs with more complete shells such as SN 1006 (Long et al. 2003), coincidence of the outer edges of radio and X-ray emission is expected. Here the displacement of the radio maximum slightly inward of the X-ray maximum may reflect the smaller extent of radio emission in the east; Figure 3 shows very similar radial extents in the western shell. The pulsar position is indicated in cyan; no radio emission is seen coincident with source region E, just to the east of the pulsar.

Most SNRs showing synchrotron X-ray emission do so in the form of "thin rims," azimuthal features at the remnant edge (Bamba et al. 2003; Vink & Laming 2003). Those rims typically have radial widths of order arcseconds, whose extents have been used to infer magnetic-field strengths, if the extents are due to radiative losses as shock-accelerated electrons convect or diffuse downstream. (See Ressler et al. (2014) and references therein.) Even the brighter regions of G32.4+0.1 are too faint for any such features to be clearly recognizable, as fairly aggressive smoothing is necessary to bring out emission above background, and no such feature is evident in the azimuthally averaged radial profiles of Figure 4.

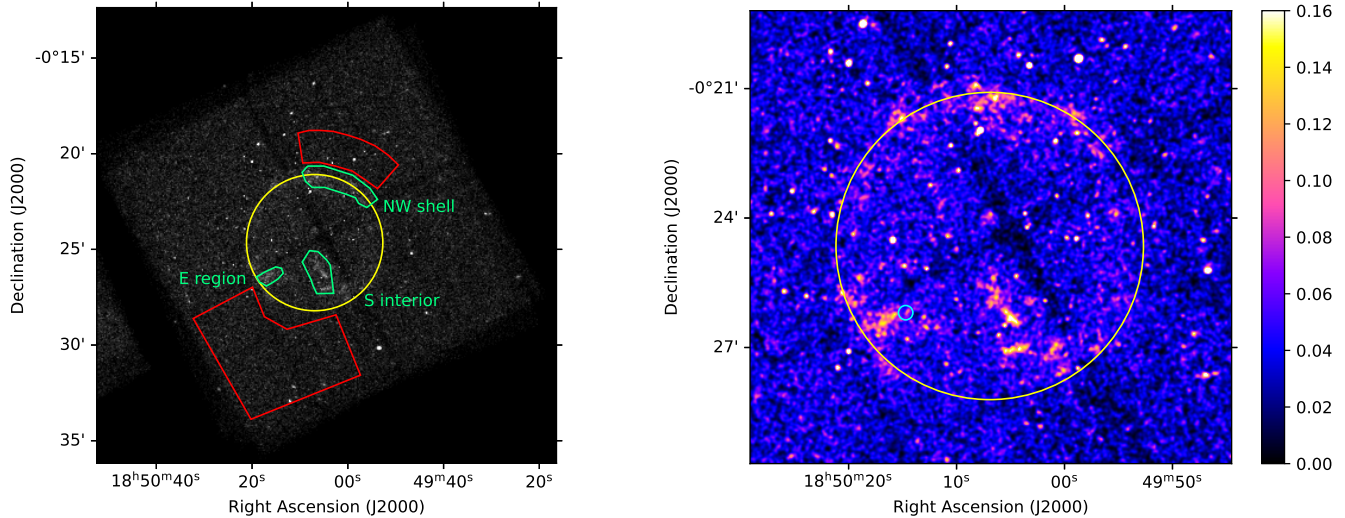

**Figure 2.** Left: Raw counts image between 1.5 and 6.5 keV, spatially binned by a factor of 4 (1.97″), indicating source regions in green and background regions in red. Upper right (NW) source region: Northwest shell, with background region C1b (for ACIS-I CCD1 chip). Lower left source region: East (E) region. Lower center (S) region: S interior. Background region C2b (for ACIS-I CCD2 chip). The yellow circle shows the adopted shell radius (210″). The image scale is linear up to 5 counts per binned pixel. Right image: G32.4+0.1 smoothed with a $\sigma = 4$ pixel (1.97″) Gaussian, between 1.5 and 6.5 keV. The cyan circle is centered on the position of the radio pulsar. The image scale is linear up to 0.16 counts per (original) ACIS pixel.

We searched for a source coincident with the radio pulsar position of RA $= 18^h\ 50^m\ (14.714\pm0.004)^s$, $\delta = -0°\ 26'\ (11.6 \pm 0.2)''$ (Keith et al. 2009). Figure 3, right, shows the vicinity of that position. No source is evident. However, ciao task srcflux reported at the pulsar position a source in the energy range of 0.5 to 7 keV with flux $5.1(0,19) \times 10^{-16}$ erg cm$^{-2}$ s$^{-1}$ ($3\sigma$ confidence interval, and uncorrected for absorption). This result assumed a power-law spectrum with photon index $\Gamma = 1.5$. Since $\Gamma$ and the absorption column $N_H$ are nearly degenerate for such a faint possible source, we simply varied $\Gamma$ between 1 and 3, producing a 50% variation on the putative source flux. We examined the region within 2″ of the pulsar position (49 pixels), finding 8 counts, 5 of which are above 3.5 keV. An annulus of radius 10″ surrounding the pulsar position contained 95 counts, predicting 3.7 within 2″ of the pulsar. Now diffuse emission in the remnant interior is undoubtedly more extensive than can easily be seen by eye, and point-source-finding algorithms can be misled by maxima in such emission. Thus we conservatively consider the flux reported by the srcflux task to be an upper limit. This possible detection will be discussed below.

## 5. SPECTROSCOPY

### 5.1. *Backgrounds*

Large background regions on each of the four ACIS-I chips have surface brightnesses ranging from 0.039 to 0.045 counts px$^{-1}$ in the 0.7–7 keV energy range, monotonically increasing from the CCD0 chip in the SW to the CCD3 chip in the NE. This suggests significant spatial variations of the X-ray sky background across the field of view. We have considered the sky background and particle background (X-rays generated by the interaction of fast particles with the spacecraft) separately. We have used the algorithm of Suzuki et al. (2021) to generate spatially dependent particle backgrounds for each source field, fit to the data at energies above those imaged by the mirrors. We then fit background fields separately on each chip with a combination of the fixed particle background for that field generated by the Suzuki et al. (2021) formalism and an absorbed power-law model to describe the remaining sky background. The latter varies by up to 40% in the 0.7–7 keV energy range. This increase is much more pronounced at low energies, by up to 90% in the 1–2 keV range, but significant variations (up to 25%) are still present in the 3–7 keV range.

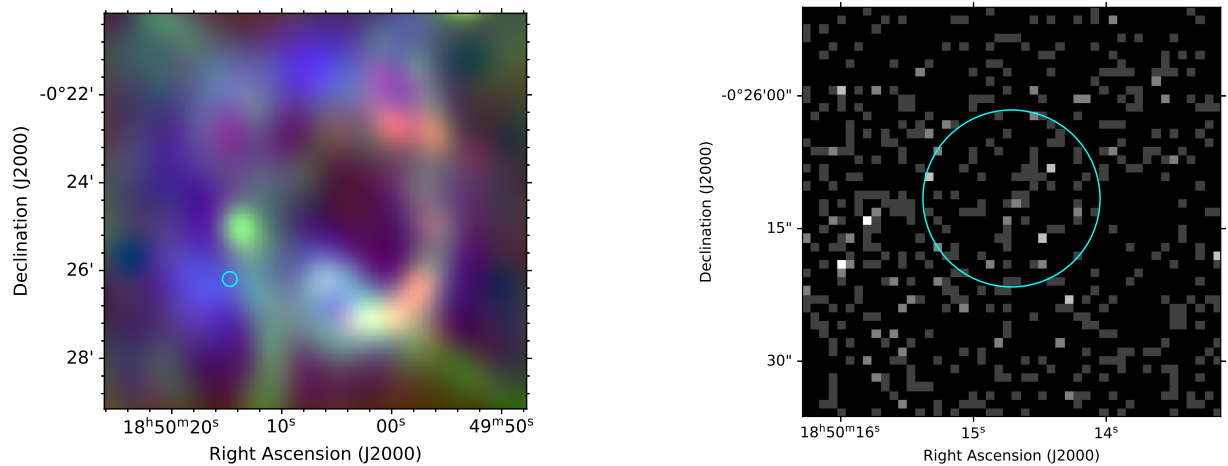

**Figure 3.** Left: RGB image of 1.4 GHz radio (NRAO VLA Sky Survey; Condon et al. 1998) in red, 5.9 GHz radio (Dokara et al. 2023) in green, and X-rays in blue, on linear scales. The X-ray image, again between 1.5 and 6.5 keV and binned by 8 as in Figure 1, has been heavily smoothed with a Gaussian with $\sigma = 9$ 3.94″ pixels to approximate the 1.4 GHz radio resolution. The 5.9 GHz radio image has been smoothed to a similar resolution. The pulsar position is indicated by the cyan circle. There appears to be no radio emission associated with the east region identified in Figure 2. Right: Close-up X-ray counts image of the radio pulsar position, binned by 2 (0.984″ pixels). The cyan circle has radius 10″. The brightness scale is linear over the entire range of 0 to 4 counts.

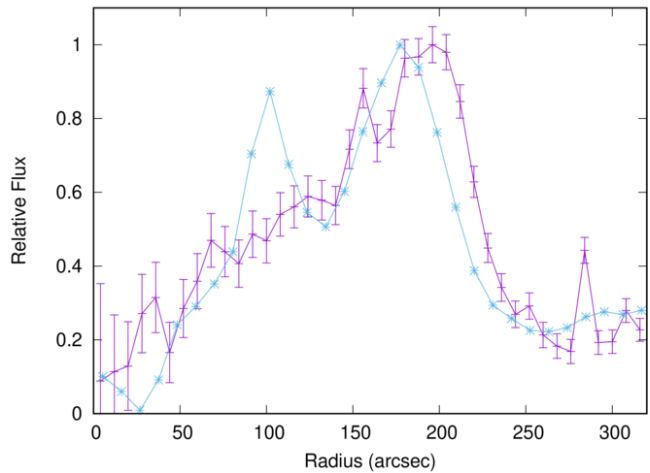

**Figure 4.** Azimuthally averaged profiles of G32.4+0.1 at 5.9 GHz (cyan, crosses) and between 1.5 and 6.5 keV (magenta plus signs, with error bars). The radio profile is obtained from the 5.9 GHz radio image shown in Figure 3 (Dokara et al. 2023). Both shells are quite irregular in azimuth, but show comparable widths when averaged. The inner peak in the radio profile at a radius of about 100″ probably results from the unrelated emission evident in Figure 3. Both profiles have been linearly rescaled to a range from 0 to 1.

There is a large-scale southwest to northeast gradient across the field of view, but more complex spatial variations are also present. This spatially-varying X-ray background is anticorrelated with local extinction (within 2–3 kpc from the Sun), as revealed by comparison with extinction maps generated using methods described by Green et al. (2019). (No reliable extinction maps with sufficiently high spatial resolution are available at larger distances.) Complex optical extinction variations across G32.4+0.1 are seen in these maps, even with the relatively poor (several arcmin) spatial resolution of these maps. X-ray sky background variations are likely to be complex as well.

Our sky backgrounds are considerably larger and more variable than the cosmic X-ray background (CXB); for the location of G32.4+0.1 in the Galactic plane within 32 degrees of the Galactic Center, we might expect contributions from the Galactic Ridge as well as other diffuse sources, and strong variations in absorption. Our background region C1b had a sky component best described by a power-law with $\Gamma = 1.54$ and $N_H = 1.6 \times 10^{21}$ cm$^{-2}$, and surface brightness between 2 and 10 keV of $3.3 \times 10^{-11}$ erg cm$^{-2}$ s$^{-1}$ deg$^{-2}$, while C2b had $\Gamma = 2.1$, $N_H$ of $1.1 \times 10^{22}$ cm$^{-2}$ and $2 - 10$ keV surface brightness of $1.9 \times 10^{-11}$ erg cm$^{-2}$ s$^{-1}$ deg$^{-2}$. Hickox & Markevitch (2006) report a CXB spectrum with $\Gamma = 1.4$ and $2 - 10$ keV surface brightness of $(1.9 \pm 0.2) \times 10^{-11}$ erg cm$^{-2}$ s$^{-1}$ deg$^{-2}$. Correcting our values for the (uncertain) absorption would raise them further above values for high-latitude CXB intensities, confirming that our sky backgrounds contain more than the CXB. In any case, we obtain our sky backgrounds by direct measurement of regions near our source regions; as long as these are well-described by simple models, we need not account in other ways for the parameter values we find.

The strong fluorescent Au M$_{\alpha,\beta}$ line complex centered at 2.1 keV is not well reproduced by the particle background models of Suzuki et al. (2021), since these models do not account for the long-term temporal gain drift of the ACIS-I CCDs. This line complex (visible in the particle background curves in Figures 5 through 7) plus the sky background dominate over the heavily-absorbed G32.4+0.1 spectra at energies less than about 2.3–2.4 keV. At these energies, spatial and spectral background variations become difficult to quantify, introducing poorly-known, systematic uncertainties that affect our spectral fits.

### 5.2. *Spectroscopic analysis*

For spectral analysis, we selected three relatively bright regions, shown in Figure 2, and restricted ourselves to the energy range $1.5 - 6.5$ keV, as the source is not apparent in images outside that range. (For the South interior region near a chip gap, we excluded OBSIDs 22430 and 25100 from the spectral extraction, as the roll angle for those two observations put the extraction region partly into the chip gap. This resulted in a loss of 22% of the exposure time for that region.) Our statistics enable us to analyze our three regions separately, which was not possible in the XMM-*Newton* study of Yamaguchi et al. (2004). The crudest but most model-independent method of describing spectra of faint sources is in terms of hardness ratios $R_H$, here defined in terms of soft and hard energy bands $S$ and $H$ as $R_H \equiv (H - S)/(H + S)$ – that is, the fractional excess of hard counts over soft. As the exposure correction should be energy-independent, we do not apply it. However, we correct for background by subtracting soft and hard background counts from indicated regions, scaled to the source region size. Results are displayed in Table 1. We note that the East region has a significantly higher (harder) ratio than the other two.

We have also performed spectral fitting with the XSPEC suite of models (Arnaud 1996), using the particle background produced by the `mkacispback` [1] routine, and including a power-law component to describe the sky background with parameters fixed to those found from fitting to a nearby background region, as described above. Given the spatially varying background, we performed fits for the three source regions shown in Figure 2 individually, and used joint fitting for combinations of regions, with each region subject to its own particle background and normalization. In all cases, the model for a region includes three components: the calculated particle-background model for the source region; the power-law model for the sky background, with fixed normalization scaled from that measured for the background region by the ratio of source to background region size; and each of three source models: a power-law, a plane shock (XSPEC model `pshock`), or a Sedov model (XSPEC model `sedov`), each including absorption. Model abundances were kept at solar values (Grevesse & Sauval 1998), as expected for a large, relatively old SNR; freeing them produced unreasonably low values ($\lesssim 0.2$) without appreciable improvement in statistical quality. For combinations of regions, all parameters except normalization were tied, except as described below. Results are shown in Table 2 and in Figures 5 through 7.

Hardness ratios suggest a harder spectrum for the East region than for the other two. To examine this possibility more closely, we performed two joint power-law fits in which all three regions shared the same absorbing column, but photon indices were allowed to vary separately. In one case, all three values were independent; in the other, we tied $\Gamma$ values in the northwest and south regions, and compared the joint value to that of the east region. These values are reported in Table 4. We find confirming evidence of a harder spectrum in the east than in the combination of the other two regions. This is illustrated in Fig. 8, where it can be seen that equal values of $\Gamma$ are disfavored with greater than 90% confidence.

## 6. DISCUSSION

---

[1] Software available at https://github.com/hiromasasuzuki/mkacispback

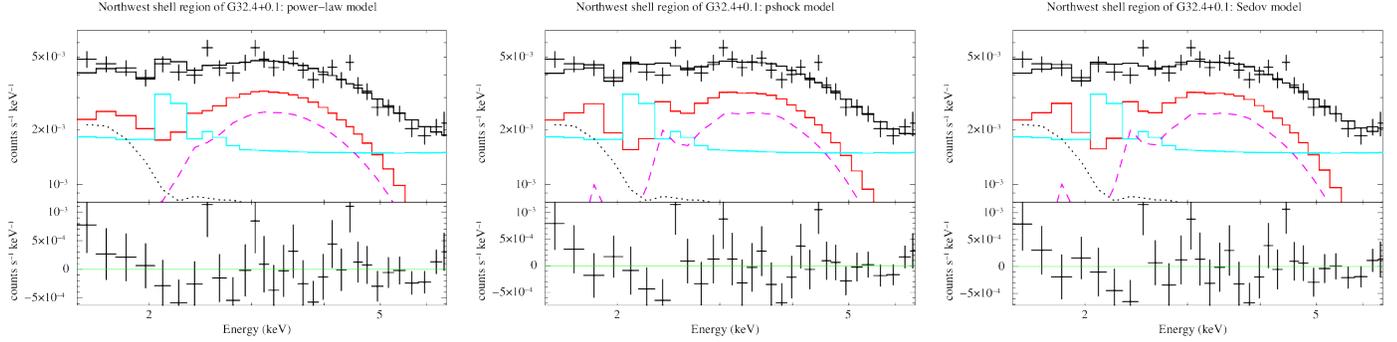

**Figure 5.** Spectra of Northwest region. Black (top) histogram: Model total along with observed data. Red (middle) histogram: source plus sky background. Magenta (dashed) curve: Source model alone (power-law). Blue (lower) histogram: particle background, calculated by the method of Suzuki et al. (2021). Dotted curve: sky background (power-law model). Parameters for the sky background were frozen to values found from analysis of the background region shown on the left panel of Figure 2, CCD1 (upper right) chip. Left: power-law fit. Center: pshock fit. Right: Sedov fit.

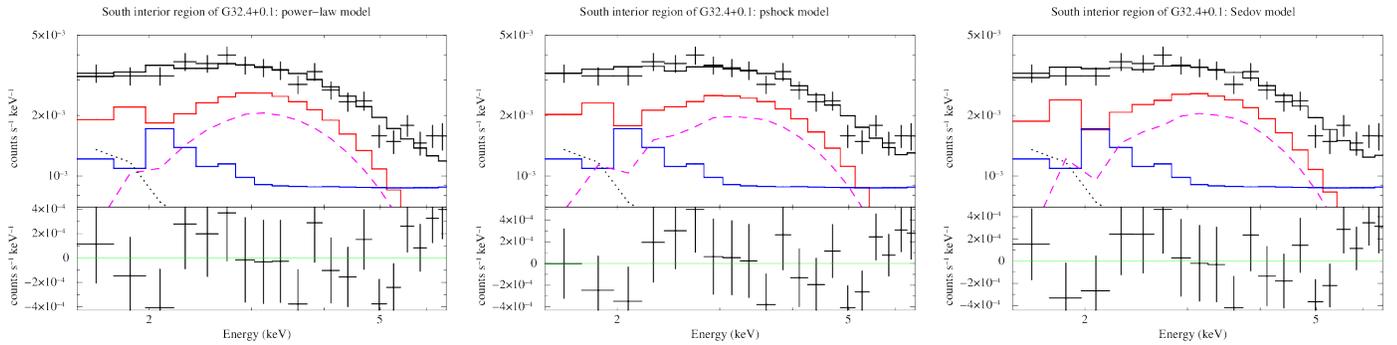

**Figure 6.** Spectra of South interior region. Lines as in Figure 5. Parameters for the sky background were frozen to values found from analysis of the background region shown on the left panel of Figure 2, CCD2 (lower left) chip. Left: power-law fit. Center: pshock fit. Right: Sedov fit.

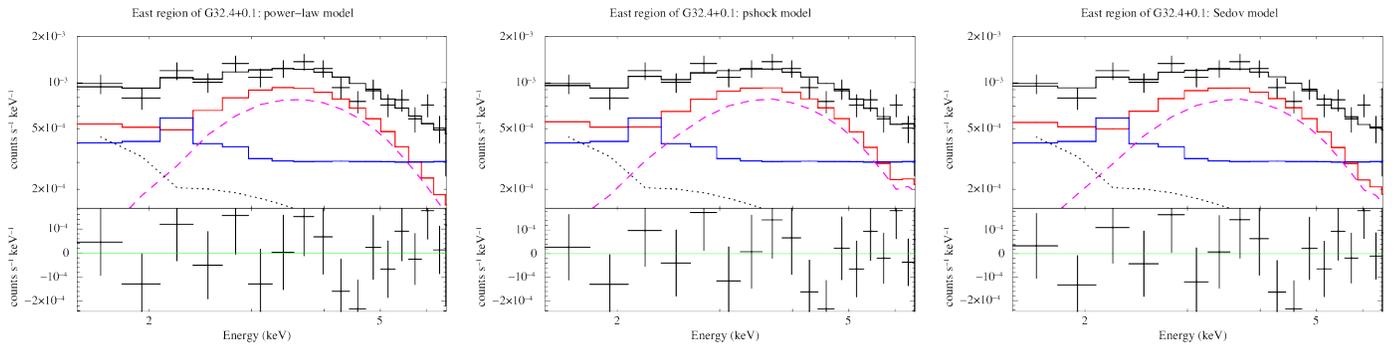

**Figure 7.** Spectra of East region. Lines as in Figure 5. Parameters for the sky background were frozen to values found from analysis of the background region shown on the left panel of Figure 2, CCD2 (lower left) chip. Left: power-law fit. Center: pshock fit. Right: Sedov fit.

### 6.1. *Supernova remnant*

G32.4+0.1 is an extremely faint SNR. Its shell, while quite variable in brightness, is fairly circular, with a mean radius of about $210''$ or 11 pc, for our assumed distance of 11 kpc. Our spectral fits are qualitatively consistent with those of Yamaguchi et al. (2004) based on XMM-*Newton* data, showing high absorption and statistically acceptable fits of power-law, plane-shock, or Sedov models. However, our results differ quantitatively somewhat from those of Yamaguchi et al. (2004); we find rather higher absorbing columns ($(6 - 11) \times 10^{22}$ cm$^{-2}$), though that is partly due to our using a different abundance set; lower temperatures ($kT \sim 1 - 4$ keV); and steeper power-law photon indices (2.7 − 4.0). Given our different treatment of background, this is perhaps not surprising. For either power-law or thermal models, however, we can deduce high shock velocities. If the X-rays are nonthermal, G32.4+0.1 is not an outlier on the $\Sigma_X - D$ diagram for X-ray synchrotron-dominated SNRs (Figure 9). The same is true for the $L_X - R$ diagram presented by Nakamura et al. (2012), though since we use 11 kpc rather than 17 kpc for the distance, the location of G32.4+0.1 on that diagram is slightly different. In either case, G32.4+0.1 fits well within the range of other large XSSNRs. Theory suggests that accelerating electrons to the TeV energies required for emission of synchrotron X-rays requires shock speeds of 1000 − 2000 km s$^{-1}$ at least (e.g., Reynolds 2008). If the emission is thermal, the electron temperatures of 1.1 − 3.7 keV shown in Table 2 imply shock velocities of 1100 − 1600 km s$^{-1}$ if electrons and ions have equal temperatures, and faster velocities if electrons and ions have not yet equilibrated, which is quite likely. In either case, then, current shock velocities in G32.4+0.1 are above 1,000 km s$^{-1}$. For generality, we write the shock velocity $v_s = mR/t$. Early phases of SNR evolution, before much mass has been swept up, can be described by self-similar solutions with $0.4 \le m \le 0.8$, given power-law profiles of SN ejecta and surrounding medium (Chevalier 1982). For Sedov evolution, $m = 2/5$. Then, for $v_s \gtrsim 1,000$ km s$^{-1}$, the remnant age $t = mR/v_s \lesssim 4,500 - 9,000$ yr.

Thermal fits then constrain the upstream density, in two ways: from the fitted values of ionization timescale $\tau = n_e t$, and from the emission measure $EM \equiv \int n_e n_H dV = \langle n \rangle^2 V$ for a homogeneous source, fixed by the flux normalization of the fits. Table 3 lists our fitted EM values from pshock fits to the three regions, along with crude estimates of the emitting volumes: just the sky-plane area to the 3/2 power. For $\tau \sim 10^{10}$ cm$^{-3}$ s and the above range of ages, $n_e \sim 0.04 - 0.07$ cm$^{-3}$. (Here we identify the SNR age $t$ with the shock age, though the latter could be less.) However, densities inferred from fitted emission measures are an order of magnitude larger, as Table 3 shows, and quite incompatible with the densities inferred from ionization timescales. The low $\tau$ values are required by the absence of distinct emission-line features. We conclude that, even though the fits are of comparable statistical quality, thermal models are not self-consistent, and we eliminate them. For a plane shock model, which sums the emission from zones with ionization timescales from zero to $\tau$ with constant electron density, the mean ionization timescale is $\tau/2$, while for a Sedov model, $\tau$ is the immediate post-shock electron density times the shock age. A fitted value of $\tau$ from the pshock model corresponds to $0.404\,\tau$(Sedov). See Borkowski et al. (2001) for more details.

Below we discuss the possibility that the East region is not part of the SNR shell, so we here consider the other two regions. The power-law fit to the combined northwest shell and south interior regions gives a photon index $\Gamma$ of 3.7 (3.1, 4.4). (Including the East region lowers the $\Gamma$ values only slightly, to 3.6 (3.0, 4.2)).This is quite steep for XSSNRS; SN 1006 has $\Gamma \sim 3$ from the nonthermally-dominated rims (e.g., Koyama et al. 1995), while the youngest Galactic SNR G1.9+0.3 has spatially varying $\Gamma$ values between 2.0 and 2.8 (Reynolds et al. 2009). A full SED model from radio to X-rays, such as srcut, which describes synchrotron emission from a power-law electron energy distribution with an exponential cutoff (Reynolds & Keohane 1999), is not well constrained due to the poor correspondence between radio and X-ray morphologies, so we have not attempted it; but the steeper spectrum in the X-rays suggests a lower maximum energy of electrons in G32.4+0.1 than in those two very young objects, which would indicate slower shocks, consistent with a greater age.

## 6.2. *Pulsar*

The radio pulsar position shows no obvious excess counts (see Figure 3, right). However, the source-finding algorithm srcflux reported, at that position, a source with observed flux between 0.5 and 7 keV of 5.1 (0,19) $\times 10^{-16}$ erg cm$^{-2}$ s$^{-1}$ for an assumed power-law index $\Gamma_{psr}$ of 1.5, typical for rotation-powered X-ray pulsars (Kargaltsev & Pavlov 2008). (For $\Gamma_{psr} = 1$, we obtain 13.7 (0,51.5) $\times 10^{-16}$ erg cm$^{-2}$ s$^{-1}$, and for $\Gamma_{psr} = 2.5$, we find 4.2 (0,16) $\times 10^{-16}$ erg cm$^{-2}$ s$^{-1}$.) We do not regard this as a significant detection, as it is consistent with zero at the $3\sigma$ level. Prinz & Becker (2015) reported a $2\sigma$ upper limit to any X-ray counterpart of the radio pulsar with a nonthermal flux between 0.1 and 2 keV of $9 \times 10^{-15}$ erg cm$^{-2}$ s$^{-1}$ based on XMM-*Newton* data. Scaled to our energy range of 0.5 − 7 keV with $\Gamma = 1.5$, this $2\sigma$ limit is $3.8 \times 10^{-15}$ erg cm$^{-2}$ s$^{-1}$, consistent with our $3\sigma$ upper limit. Shibata et al. (2016) collect data on X-ray luminosities of rotation-powered pulsars. According to their Figure 1, our very conservative upper limit is near the top of the distribution of pulsars with $\dot{E} \sim 10^{35.5}$ erg s$^{-1}$ – that is, it is unconstraining.

At present, the only evidence for an association of the pulsar and the SNR (in addition to position) is the large, but uncertain, column density toward the SNR. The age estimate for G32.4+0.1 obtained above is much less than the characteristic spindown age $t_{ch} \equiv P/2\dot{P}$ of 68 kyr. (A SNR this old, with a current shock velocity of 1000 km s$^{-1}$, would require a combination of supernova explosion energy $E_{51} \equiv E_0/10^{51}$ erg and upstream density of $E_{51}/n_0 \sim 10^4$ erg cm$^3$ – highly implausible.) If we assume the pulsar was born in the SN event only a few thousand years ago, then (in the normal pulsar formalism) the pulsar must have been born with a period not much shorter than its current 166 ms. A pulsar losing energy due to magnetic dipole radiation with a constant braking index $n \equiv P\ddot{P}/\dot{P}^2$ has a period and a luminosity $L_{\rm tot}$ (total energy-loss rate) that vary with time as

$$P = P_0\,(1 + t/\tau_c)^{1/(n-1)} \quad \text{and} \quad L_{\rm tot} = L_0/\,(1 + t/\tau_c)^{(n+1/n-1)} \tag{1}$$

where $\tau_c \equiv 2t_{\rm ch}/(n-1) - t$ and $t$ is the true age. Assuming the braking index has been constant at 3 for the life of the pulsar, and taking 5,000 years for the true age, we have $\tau_c = 63,000$ yr, so $t/\tau_c = 0.079$, $P_0 = 160$ ms, and $L_0 = 3.8 \times 10^{35}$ erg s$^{-1}$ – that is, both almost unchanged since birth. The total spindown energy lost so far is then about $L_0\,t \sim 6.0 \times 10^{46}$ erg; the total rotational energy at birth is about $L_0\,\tau_c \sim 8.2 \times 10^{47}$ erg, not an unreasonable value.

A more serious objection to an association, or to the simple SNR age estimate above, is the large offset of the pulsar from the location of the supernova, which we reasonably assume to be near the center of symmetry of the fairly round shell. That offset, about $150''$, corresponds at 11 kpc to a distance of about 8 pc, and would require a sky-plane velocity of about $1600\,d_{11}(t/5000 \text{ yr})^{-1}$ km s$^{-1}$ – highly unlikely. This result calls into question the SNR age or the association. (The pulsar distance, based on the observed dispersion measure, is likely the most reliable.) The SNR age estimate is based on interpreting the X-ray spectrum as synchrotron emission. Assuming a minimum shock velocity for accelerating electrons to the required TeV energies means that the inferred remnant age is proportional to distance, so that the required pulsar velocity for a given angular offset is independent of distance.

If the pulsar is not associated with the shell, the agreement in both position and (roughly) in distance is fortuitous, and the pulsar is truly much older than the SNR, allowing its initial period $P_0$ to be significantly shorter than the current 166 ms. Additionally, if the East region is really a PWN produced by the pulsar (see below), it could be interpreted as a bow-shock nebula trailing behind the pulsar moving to the west (toward, rather than away from, the shell center).

### 6.3. Pulsar-wind nebula?

The East region (see Figure 2) extends to the east from the nominal pulsar position. It is possible that this extended emission is related to the pulsar. The hardness ratios of Table 1 show that this region has a considerably harder spectrum than the other source regions, a conclusion supported by the spectral fitting described above and illustrated in Figure 8. We conclude that it is moderately likely that the East region is in fact a pulsar-wind nebula associated with PSR J1850-0026. (This association is independent of whether the pulsar is related to the SNR.) It is important to note, however, that our best-fit photon index $\Gamma_{\rm E}$ of 2.7 would be the second steepest spectrum of the 54 PWNe catalogued by Kargaltsev & Pavlov (2008), including 14 objects with no known pulsar, but the uncertainties in our fit are large.

We obtained from the power-law fit to the E region described in Table 2 an observed (i.e., absorbed) flux between 0.5 and 8 keV of $7.8(6.9, 8.8) \times 10^{-14}$ erg cm$^{-2}$ s$^{-1}$, giving for a distance of 11 kpc a luminosity $L_x$ of $1.1\,(1.0, 1.3) \times 10^{33}$ erg s$^{-1}$. Correcting for the (very large) absorption raises these values by a factor of about 2, to $S_{\rm PWN} = 1.7\,(1.4, 1.9) \times 10^{-13}$ erg cm$^{-2}$ s$^{-1}$ and $L_{\rm PWN} = 2.3\,(2.1, 2.8) \times 10^{33}$ erg s$^{-1}$. This luminosity would make the PWN considerably more luminous for the pulsar $\dot{E}$ than typical (Figure 5 of Kargaltsev et al. 2008), with a value of $L_x/\dot{E}$ of $7 \times 10^{-3}$, near the empirical upper-bound relation of Kargaltsev et al. (2013).

The 5.9 GHz radio image shown in Dokara et al. (2023) shows no obvious radio counterpart (see Figure 3). The E region contains about 2.2 mJy (= 3.3 erg erg$^{-1}$ cm$^{-2}$ s$^{-1}$), but this value is comparable to that found in other regions of the same size away from obvious emission in the image, so we regard it as an upper limit. The normalization of the power-law fit to the E region X-ray emission is $1.7 \times 10^{-4}$ photons keV$^{-1}$ cm$^{-2}$ s$^{-1}$ at 1 keV, or $1.7 \times 10^{-4}$ erg erg$^{-1}$ cm$^{-2}$ s$^{-1}$. These values bound the mean radio-to-X-ray energy spectral index $\alpha_{rx}$ from below at $-0.56$, i.e., any radio counterpart must have a flatter radio-to-X-ray spectral-energy distribution than this value. This corresponds to a limit on photon index $\Gamma_{rx}$ of $\Gamma_{rx} < 1.56$, a much flatter spectrum than the X-ray power-law. This is not surprising, as no PWN has a value of $\Gamma_{rx}$ as large (steep) as its X-ray value $\Gamma_x$. In general, if the East region is a pulsar-wind nebula, it is an unusual one. Unfortunately, the faintness of the source means that confirmation will be difficult.

**Table 1.** Hardness ratios for different regions

| Region | Background region | S0 | SC | H0 | HC | HR |
|---|---|---|---|---|---|---|
| Northwest shell | C1b | $1613 \pm 40$ | $605 \pm 51$ | $1599 \pm 40$ | $687 \pm 50$ | $0.064 \pm 0.036$ |
| East region | C2b | $380 \pm 20$ | $168 \pm 24$ | $455 \pm 21$ | $263 \pm 25$ | $0.22 \pm 0.083$ |
| South interior | C2b | $1136 \pm 34$ | $505 \pm 42$ | $1049 \pm 32$ | $478 \pm 40$ | $-0.027 \pm 0.040$ |
| Background Chip 1b (C1b) | | $2180 \pm 47$ | | $1971 \pm 44$ | | $-0.05$ |
| Background Chip 2b (C2b) | | $8383 \pm 92$ | | $7580 \pm 87$ | | $-0.05$ |

NOTE—S0: raw soft counts ($1.5 - 3.5$ keV); SC: soft counts corrected for background; H0: raw hard counts ($3.5 - 6.5$ keV); HC: background-corrected hard counts; HR, hardness ratio (HC − SC)/(HC + SC). Uncertainties are $\sqrt{N}$, combined in quadrature.

**Table 2.** Spectral fits to different regions

| Region | Source model | $N_{\rm H}(\times 10^{22})$ cm$^{-2}$ | $\Gamma$ or $kT$ | $\tau \equiv n_e t$ (shock models) | cstat/dof |
|---|---|---|---|---|---|
| Northwest shell | `powerlaw` | 11 (8.3, 15) | 4.0 (3.2, 5.1) | | 338.8/342 |
| Northwest shell | `pshock` | 11 (8.4, 14) | 1.5 (1.1, 2.1) | 1.1 (0, 2.6) $\times 10^{10}$ cm$^{-3}$ s | 336.6/342 |
| Northwest shell | `sedov` | 11 (8.3, 16) | 1.1 (0.55, 1.5) | 2.2 (0, 4.8) $\times 10^{10}$ cm$^{-3}$ s | 336.3/342 |
| South interior | `powerlaw` | 8.2 (6.2, 11) | 3.4 (2.7, 4.2) | | 384.5/342 |
| South interior | `pshock` | 6.4 (5.2, 8.7) | 2.6 (1.8, 3.7) | 4.9 (0, 92) $\times 10^{8}$ cm$^{-3}$ s | 381.2/342 |
| South interior | `sedov` | 7.1 (4.9, 9.5) | 1.6 (0.90, 2.8) | 1.2 (0, 2.6) $\times 10^{10}$ cm$^{-3}$ s | 378.7/342 |
| East region | `powerlaw` | 9.4 (5.6, 15) | 2.7 (1.6, 4.1) | | 369.4/342 |
| East region | `pshock` | 8.3 (6.0, 12) | 3.7 (1.9, 8.1) | 1.0 (0, 100) $\times 10^{8}$ cm$^{-3}$ s | 369.4/342 |
| East region | `sedov` | 8.2 (5.6, 13) | 2.9 (1.2, $\infty$) | 1.2 (0, 54) $\times 10^{9}$ cm$^{-3}$ s | 369.6/342 |
| Combined | `powerlaw` | 9.6 (7.8, 12) | 3.6 (3.0, 4.2) | | 1099/1020 |
| Combined | `pshock` | 7.8 (6.6, 11) | 2.2 (1.7, 2.9) | 4.1(0, 110) $\times 10^{8}$ | 1099/1019 |
| Combined | `sedov` | 9.4 (7.0, 12) | 1.3 (0.94, 2.1) | 1.7(0, 3.0) $\times 10^{10}$ | 1097/1019 |

NOTE—Quoted bounds are 90% confidence intervals. All fits are between 1.5 and 6.5 keV. For Sedov models, temperatures are electron temperatures, with mean temperatures frozen to the mean value for a shock velocity of 2000 km s$^{-1}$, $kT_s = 4.6$ keV.

**Table 3.** Emission measures from `pshock` spectral fits

| Region | Sky-plane area (arcsec$^2$) | Volume (cm$^3$) | EM (cm$^{-3}$) | $\langle n \rangle$ (cm$^{-3}$) | $\tau$ (cm$^{-3}$ s) | $\tau/\langle n \rangle$ (yr) |
|---|---|---|---|---|---|---|
| Northwest shell | 14,100 | $7.5 \times 10^{57}$ | $3.8 \times 10^{57}$ | 0.71 | $1.1 \times 10^{10}$ | 490 |
| South interior | 8650 | $3.6 \times 10^{57}$ | $9.5 \times 10^{56}$ | 0.51 | $4.9 \times 10^{8}$ | 30 |
| East region | 2800 | $6.7 \times 10^{56}$ | $3.2 \times 10^{56}$ | 0.69 | $1.2 \times 10^{9}$ | 55 |

NOTE—Volumes were estimated as (sky area)$^{3/2}$. A distance of 11 kpc was assumed, so that 1 arcsec$^3 = 4.49 \times 10^{51}$ cm$^3$.

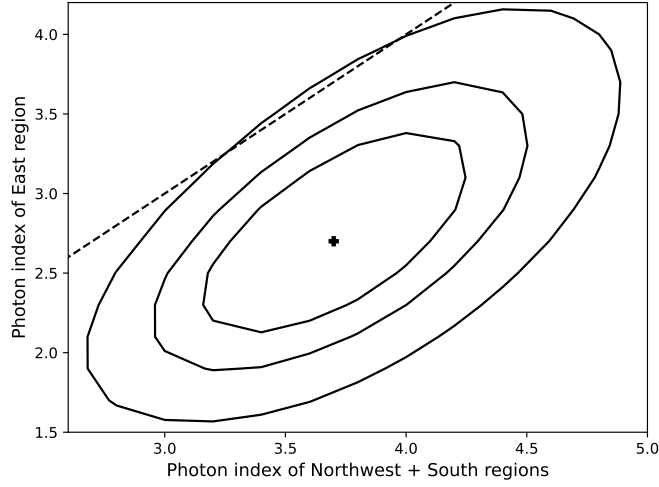

**Figure 8.** Contour plot of C-statistics for photon indices $\Gamma$ resulting from joint fits of a power-law model for combined regions Northwest and South ($\Gamma_{NS}$), compared with East ($\Gamma_E$). Contours are $\Delta$Cstat = 2.3, 4.6, and 9.2 (68%, 90%, and 99%) with respect to the best-fit value (central cross). The line $\Gamma_{NS} = \Gamma_E$, shown dotted, lies at the edge of the line $\Delta$Cstat = 9.2.

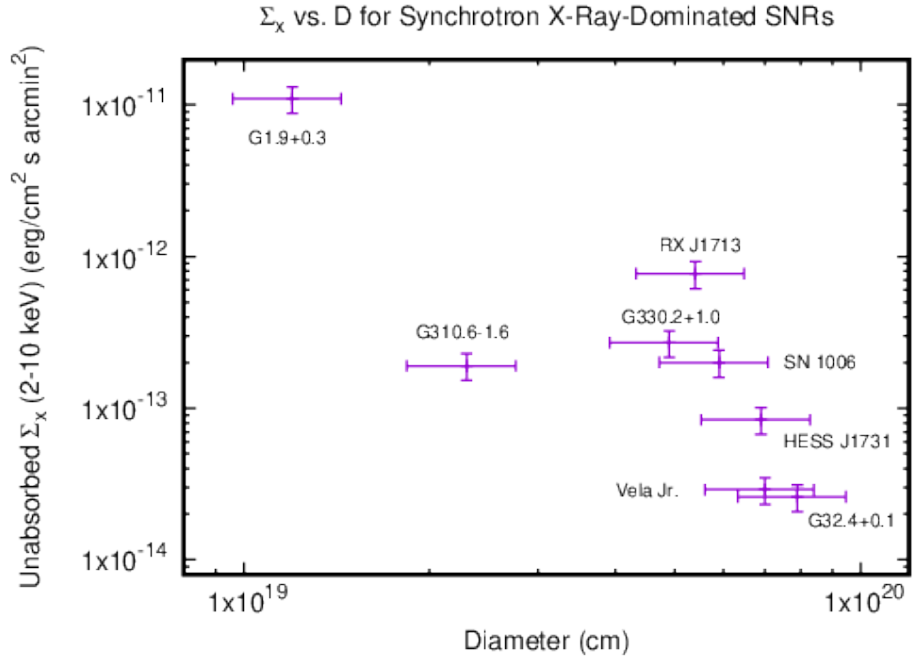

**Figure 9.** X-ray surface brightness-diameter plot for eight Galactic synchrotron X-ray dominated supernova remnants (Reynolds & Borkowski 2019, and references therein). For the larger, fainter objects, low surface-brightness emission may have been missed. For G32.4+0.1, the flux is assumed to be the sum of the fluxes of the three bright regions.

## 7. CONCLUSIONS

**Table 4.** Joint power-law fits to three regions

| Fit | $N_{\mathrm{H}}$ | $\Gamma$(South) | $\Gamma$ (Northwest) | $\Gamma$ (East) | cstat/dof |
|---|---|---|---|---|---|
| Separate photon indices | 9.5 (7.8, 12) | 3.9 (3.2, 4.7) | 3.6 (2.9, 4.2) | 2.8 (2.1, 3.5) | 1089/1018 |
| Northwest and South tied | 9.4 (7.7, 11) | 3.7 (3.1, 4.4) | $\equiv \Gamma$ (South) | 2.7 (2.0, 3.5) | 1091/1019 |
| All three tied | 9.6 (7.8, 12) | 3.6 (3.0, 4.2) | $\equiv \Gamma$ (South) | $\equiv \Gamma$ (South) | 1099/1020 |

NOTE—Only normalizations were allowed to vary, in addition to the power-law indices and absorption as indicated.

We conclude that G32.4+0.1 is a bona fide member of the class of X-ray-synchrotron dominated SNRs, since the absence of line features leads to inconsistencies in thermal models. Its large size and low surface brightness make it one of the most extreme of the class. We infer an age of less than about 9,000 yr, and surrounding densities of 0.04 – 0.07 cm$^{-3}$, implying low post-shock pressures and low energy densities in nonthermal particles and magnetic field, thereby explaining the low surface brightness of G32.4+0.1.

We did not definitively detect the pulsar in X-rays. Its large offset from the center of the SNR makes an association unlikely. While the pulsar's true age could be much less than the spindown age of 68,000 yr if it was born with a period not much shorter than currently observed, if it was produced in the supernova that resulted in G32.4+0.1, a typical pulsar kick velocity of 300 km s$^{-1}$ would give an age of 37,000 years and a shock velocity far too slow to accelerate particles to TeV energies.

The extended emission region just east of the pulsar position has a harder spectrum than shell emission, and could be a pulsar-wind nebula produced by this pulsar, most plausibly if it is a bow-shock nebula with the pulsar at the apex, requiring pulsar motion to the west and firmly ruling out an association with the SNR. It could also be a patch of SNR emission, with no X-rays associated with the pulsar or with any pulsar-wind nebula. Its high absorption is consistent with that from the SNR, suggesting some kind of association.

G32.4+0.1 may represent the lower surface-brightness limit for unambiguous identification of highly absorbed and distant XSSNRs in the Galactic plane. The combination of high absorption and high and variable background requires much deeper observations with *Chandra* or XMM-*Newton* to yield much more substantial information. More detailed radio studies may provide the best source of new information about this complex object.